# Smart Diagnosis and Early Intervention in PCOS: A Deep Learning Approach to Women's Reproductive Health

Shayan Abrar
*Dept. of CS*
*American International University-Bangladesh*
Dhaka, Bangladesh
shayanabrar7@gmail.com

Samura Rahman
*Dept. of CSE*
*East West University*
Dhaka, Bangladesh
samura.rahman14@gmail.com

Ishrat Jahan Momo
*Dept. of CSE*
*East West University*
Dhaka, Bangladesh
ishratjahanmomo13@gmail.com

Mahjabin Tasnim Samiha
*Dept. of CSE*
*East West University*
Dhaka, Bangladesh
tasnimsamiha547@gmail.com

B. M. Shahria Alam
*Dept. of CSE*
*East West University*
Dhaka, Bangladesh
bmshahria@gmail.com

Mohammad Tahmid Noor
*Dept. of CSE*
*East West University*
Dhaka, Bangladesh
tahmidnoor770@gmail.com

Nishat Tasnim Niloy
*Dept. of CSE*
*East West University*
Dhaka, Bangladesh
nishat.niloy@ewubd.edu

*Abstract*— Polycystic Ovary Syndrome (PCOS) is a widespread disorder in women of reproductive age, characterized by a hormonal imbalance, irregular periods, and multiple ovarian cysts. Infertility, metabolic syndrome, and cardiovascular risks are long-term complications that make early detection essential. In this paper, we design a powerful framework based on transfer learning utilizing DenseNet201 and ResNet50 for classifying ovarian ultrasound images. The model was trained on an online dataset containing 3856 ultrasound images of cyst-infected and non-infected patients. Each ultrasound frame was resized to 224×224 pixels and encoded with precise pathological indicators. The MixUp and CutMix augmentation strategies were used to improve generalization, yielding a peak validation accuracy of 99.80% by Densenet201 and a validation loss of 0.617 with alpha values of 0.25 and 0.4, respectively. We evaluated the model's interpretability using leading Explainable AI (XAI) approaches such as SHAP, Grad-CAM, and LIME, reasoning with and presenting explicit visual reasons for the model's behaviors, therefore increasing the model's transparency. This study proposes an automated system for medical picture diagnosis that may be used effectively and confidently in clinical practice.

## I. Introduction

Polycystic Ovary Syndrome (PCOS) is one of the most common hormonal disorders. Affecting approximately 8% to 13% of women of reproductive age Worldwide. Depending on the diagnostic criteria used. The prevalence may rise to nearly 20%. In countries like India, studies have shown that the prevalence among young ladies is approximately 9.13%. Reaching over 18% in urban areas. PCOS is a major contributor to female infertility. accounting for around 70% of ovulation-related infertility cases. Clinically, it is specified by the presence of multiple ovarian cysts, chronic anovulation, and hyperandrogenism. Despite its global health suggestion. PCOS often goes undiagnosed or is detected too late due to symptom variability and the reliance on dependence evaluation of ultrasound and hormonal data.

Recent research has used deep learning techniques to automate PCOS detection and improve diagnostic accuracy. In their study, Shankar achieved 88.4% accuracy using a CNN model applied to scleral images. Introducing a novel noninvasive diagnostic approach [1]. Similarly utilized convolutional neural networks on ultrasound images, achieving 95.4% accuracy. Their work demonstrated that CNN-based models could effectively extract relevant features for early PCOS diagnosis and emphasized the importance of proper image preprocessing and model tuning [2].

We chose to work on PCOS because, despite its high prevalence and serious health issues. It often remains underdiagnosed or misdiagnosed. particularly in developing regions. For years, many women have suffered from symptoms including infertility, weight gain, pimples, and irregular menstruation without getting the proper medical care. Lack of knowledge, restricted access to medical equipment, and the challenge of manually interpreting ultrasound pictures are often to blame for this delay. Traditional diagnostic practices typically depend on hormonal testing and expert radiological assessment. Both of which need particular equipment and a qualified person, which are not usually available in rural areas or areas with limited resources. By developing an automated, image-based detection system using deep learning. Our goal is to reduce human error, cut off diagnostic delays, and provide reasonably priced screening technology that can be implemented in poor regions and clinics.

The main goal of this research is to build a non-invasive, scalable, and accurate PCOS detection system that can support healthcare professionals, particularly in resource-constrained settings. By automating the detection process using ultrasound imaging and deep learning. We aim to facilitate early diagnosis and timely treatment, ultimately improving patient outcomes.

Research Questions:
- Can deep convolutional neural networks effectively classify PCOS using ultrasound images with high accuracy?
- What impact does image preprocessing have on the model's ability to detect PCOS-relevant features?

- What level of explainability can be achieved through visual tools like Grad-CAM to assist physicians in validating model predictions?

## II. RELATED WORKS

Over the past few years, with the help of artificial intelligence, researchers have now developing tools that make it easier to diagnose PCOS early and more accurately. Several recent studies show how deep learning, hybrid machine learning techniques, and smart image analysis are improving healthcare in this field.

Kumar et al. [3] focused on image preprocessing and they came out with a segmentation technique that mixes fuzzy C-means clustering with milestone algorithms to enhance the validation of ovarian features in ultrasound scans. This allowed for a more accurate rate of indices like follicle count and size. This led to a final classification accuracy of 92.8%. Their work shows the importance of accurate segmentation before putting the data into any classification model.

Zad et al. [4] used a large-scale electronic health record (EHR) dataset. More than 30,000 women to build machine learning based models to detect polycystic ovarian syndrome (PCOS). Among the four diagnostic criteria groups. The model's area under the curve (AUC) values were 85%, 81%, 80%, and 82%. To correct the class imbalance, weighted F1-scores have been used..

An ideal ensemble learning framework (WaOEL) has been proposed by Panjwani et al. [5] for the early detection of PCOS using common symptoms and behavioral data. By combining the heart disease and PCOS datasets. They created a new mixed dataset and trained multiple classifiers. WaOEL scored the best of the optimized ensemble models with 92.8% accuracy, 0.93 AUC, and 0.91 F1 score..

One big approach was introduced by W. Lv et al. [1] Who used eye scans as a noninvasive diagnostic alternative With these eye images their convolutional neural network (CNN) model was trained to detect signs of hormonal imbalance with these eye images. Achieving an accuracy of 88.4%. This innovative method highlights how accessible tools like eye imaging could play a role in early PCOS screening. Especially in resource-limited settings where internal imaging might not be readily available.

Sumathi et al. [2] focused on analyzing ultrasound images using CNNs. Their model was specifically trained on grayscale ovarian scans. The preprocessing steps are designed to increase the clarity of hard visual features. by using methods that made follicular structures clearer to see. They succeeded in reaching 95.4% accuracy. Their research shows how important high-quality input images are to the accurate detection of minor clinical patterns by deep learning models.

Together, these studies paint a promising picture for AI in reproductive health. Researchers are building systems that make PCOS diagnosis more accurate, accessible, and reliable. These advancements hold great promise.

## III. METHODOLOGY

The Fig. 1 of this study is a methodology diagram. The dataset was systematically loaded, and then treatment of duplicates and missing values to get clean data was done rigorously. Appropriate categorical attributes were encoded, and outliers were carefully removed to ensure robustness of the dataset. Scaling procedures were used to standardize the data distribution and assess the feature significance. Then, the processed dataset was partitioned into a training set and a testing set. To improve model efficiency, feature selection techniques were applied. Then, multiple models were trained independently, and their fusion was completed strategically, selecting the best–performing fusion model.

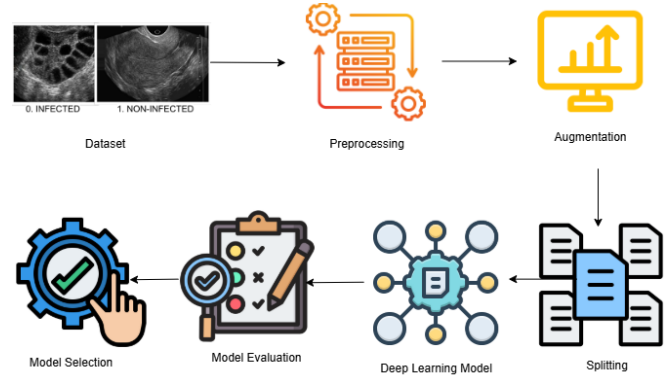

Fig. 1. Methodology

The dataset consisted of 3,840 ultrasound images divided into two classes: infected and non-infected [6]. Steps were taken within the dataset to improve its readability and fit with the model in the best possible way. Each image was resized to 224×224 pixels and then rescaled, with center-cropping performed to maintain spatial consistency. The value of the pixels in the entire dataset image was brought to the same level by performing normalization. Fig. 2 shows samples from the dataset that portray the diversity and clarity needed for effective training of the models.

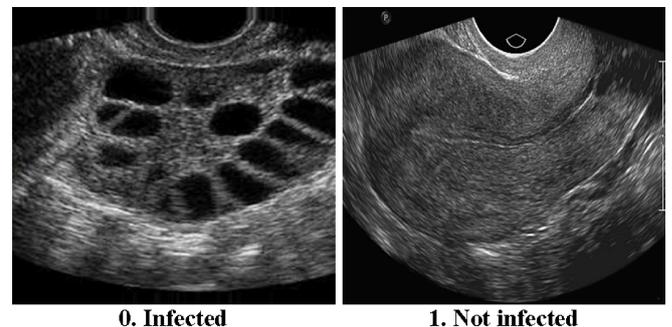

Fig. 2. Dataset Sample

### A. DenseNet201

The deep convolutional neural network known as DenseNet201 is made up of 201 layers, and it presents a novel connection scheme in which every layer is connected to all the other succeeding layers. Compared to other architectures, DenseNet201 improves the likelihood of the vanishing gradient problem happening, increases parameter reduction, and feature reuse. Due to each layer receiving feature maps from the preceding layers, the feature propagation is very effective and efficient. The optimized structure of DenseNet201 aids it in achieving improvements in performance and optimizing the speed of convergence and generalization across multi-class classification tasks, which are complex.

Fig. 3 illustrates the DenseNet201 architecture, which has dense connectivity between layers to improve gradient flow and feature reuse. In the traditional CNN architectures, information flows sequentially, but DenseNet201 connects

IEEE - 6483316th ICCCNT IEEE Conference,
July 6 - 11, 2025,
IIT - Indore, Madhya Pradesh, India

each layer to all subsequent layers. This model consists of multiple dense blocks, where each layer receives inputs from all previous layers, which enhances learning efficiency while reducing the parameters. To help extract features, the batch normalization, ReLU, and Convolution layers are used, while fully connected layers with dropout refine the classification. This design makes DenseNet201 suitable for PCOS classification.

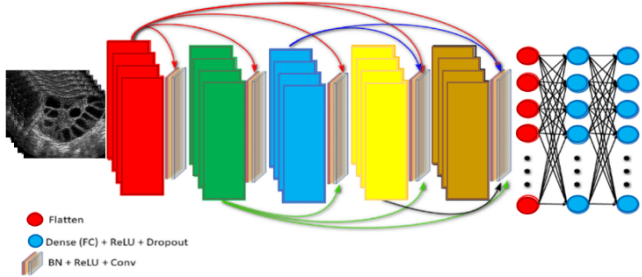

Fig. 3. Proposed DenseNet201 architecture

### B. ResNet50

ResNet50 is a very deep convolutional neural network of 50 layers and was revolutionary in solving the challenges in training very deep networks. ResNet50 learns a residual mapping instead of the underlying mapping, thereby easing the problem of training very deep networks and preventing the degradation problem [7]. ResNet50 learns features with an effective hierarchical order when the residual blocks are balanced with batch normalization and identity mappings.

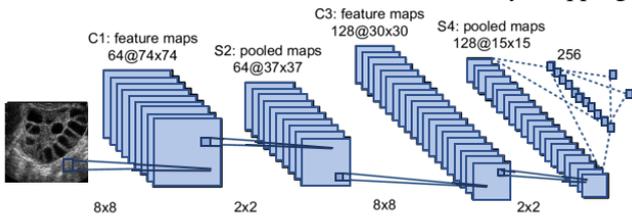

Fig. 4. Proposed ResNet50 architecture

Fig. 4, shown above, illustrates a typical convolutional neural network feature extraction pipeline represented by ResNet50. It starts with the input image passing through convolution (C1: 64@74×74 with 8×8 filters), pooling (S2: 64@37×37 with 2×2 pooling), convolution (C3: 128@30×30 with 8×8 filters), and pooling (S4: 128@15×15 with 2×2 pooling), finishing with a final fully connected layer of 256 units. The hierarchical structure reflects the multi-level abstraction and discriminative power embedded in ResNet50's architecture.

## IV. AUGMENTED REGULARIZATION

### A. CutMix

The CutMix technique which is shown in Fig. 5 is based on the augmentation of the original ultrasound image with a sample from another image in the test set, but this time the shape of the substitution region is a patch of pixels and, importantly, the substitution includes both the source image and the source label of the patch. This introduces spatial diversity and thus helps generalization by introducing such variations. The patch area is used to adjust the target label so that the learning between image segments is balanced. For the input and label denoted as x and y, CutMix is defined as:

$$xx\_new = M \odot x\_i + (1-M) \odot x\_j$$
$$y\_new = \partial y\_i + (1-\partial) y\_j$$

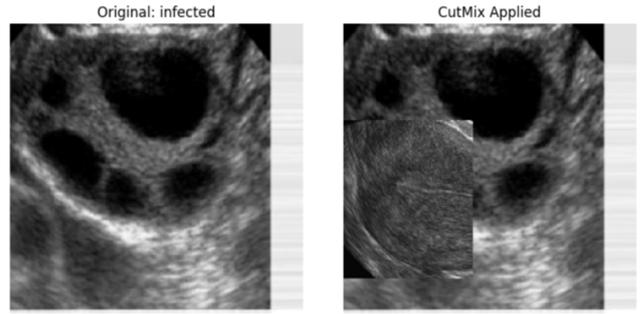

Fig. 5. CutMix image for infected = 0.3

### B. MixUp

Fig. 6, MixUp is defined as linearly blending two ultrasound images by interpolating their pixel intensities as well as their corresponding labels. For randomly chosen input pairs, xi and xj & yi and yj. The mixing thus yields a convex combination of the original mixed image and label. This approach yields smooth transitions between classes, alleviates overfitting, and strengthens model robustness by hopefully making the network act linearly in between training datapoints. MixUp can be defined as:

$$x\_new = \partial x\_i + (1-\partial) x\_j$$
$$y\_new = \partial y\_i + (1-\partial) y\_j$$

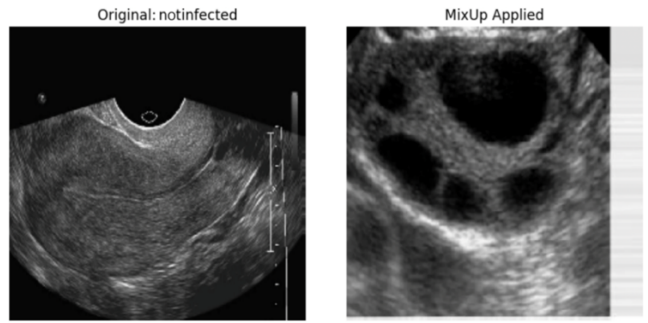

Fig. 6. MixUp image for notinfected = 0.2

## V. EXPERIMENTAL ANALYSIS

### A. DenseNet201

In Fig. 7, the training and validation accuracy curves and training and validation loss for DenseNet201 show high performance for 98 epochs. Looking at the model from the first epoch, the training and validation accuracy both exceed 98% at an early stage in the training process. The curves are still very tightly coupled, which implies little overfitting and strong generalization power. This is despite minor fluctuations, as seen particularly in a very visible dip in validation accuracy around epoch 40, but this seems to recover sooner than later and settles at almost perfect (~100%) accuracy on training and validation sets. It may not harm the model at all, since it is a temporary dip that can be caused by batch variance or noise in the validation set. The upward trajectory seems smooth and plateaus, indicating that DenseNet201 can discover discriminative features that are useful for accurate classification, and applies to the PCOS ultrasound image detection tasks.

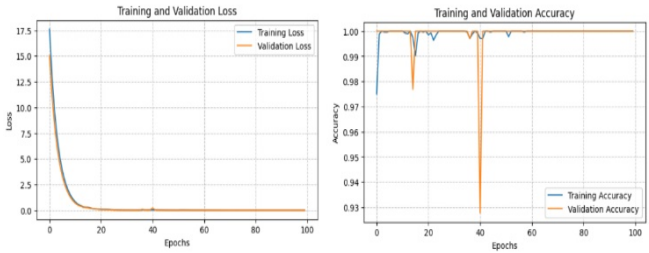

Fig. 7. a) Accuracy for training and validation. b) Loss for training and validation

The test dataset shows an impeccable classification performance of the DenseNet201 model and the confusion matrix. In total, all 162 infected cases and 223 non-infected cases were accurately picked. The difference in models is perfect bifurcation between the two classes, which indicates the exceptional discriminative power and class-specific feature capability of the model. Such a flawless result indicates that a tradeoff between sensitivity and specificity is very good, with precision, recall, and F1 score all attaining their theoretical maximums. As a result, this model is robust and reliable due to its capacity to generalize well across both classes, which makes it a viable candidate for use in real world automated PCOS detection systems. The DenseNet201 model can also produce clinically meaningful results with very high fidelity, which is reaffirmed by using the confusion matrix shown in Fig. 8.

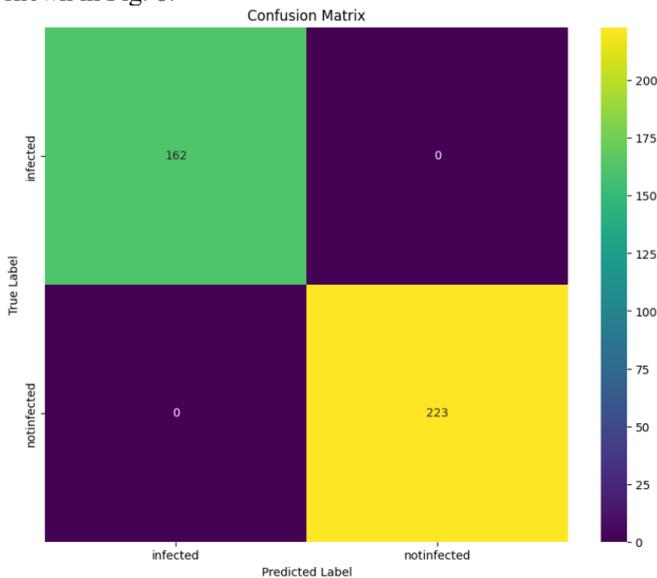

Fig. 8. Confusion Matrix of DenseNet201

Assessing the performance of the DenseNet201 model for classifying PCOS shows excellence with 100% test accuracy and a test loss of 4.54e-05, as well as flawless precision, recall, and F1-scores for both infected and non-infected classes. The training accuracy was 99%, and the validation accuracy was 100%, reflecting strong overfitting. Other hyperparameters included 100 epochs, a batch size of 32, using the Adam optimizer with a 0.0001 learning rate, binary cross entropy, and early stopping set at 15. The model was trained with data Augmentation and checkpointing alongside improving validation loss, demonstrated through saved model checkpoints. The model's fidelity was further substantiated through a confusion matrix and classification reports proving its effectiveness for PCOS detection. Each hyperparameter used for the architecture DenseNet201 in our research is shown in Table 1.

TABLE I HYPERPARAMETER TUNING OF DENSENET201

| Batch size | 32 | Loss function | binary_crossentropy |
|---|---|---|---|
| Learning rate | 0.0001 | Number of epochs | 98 |
| Optimizer | Adam | Patience | 15 |

### B. ResNet50

The convergence plot of the overall performance of ResNet50 across 67 epochs is depicted in Fig. 9. Training debiases the model. There is a minimal gap between training accuracy and validation accuracy starting from the very beginning of the training phase. Despite the presence of some training/validation accuracy plateaus, particularly in the early and middle of the training process, the model proves that it can recover, indicating some robustness to data perturbations. These accuracy trends are consistent with the loss curves, where we observe sharp declines in both training and validation loss, followed by stabilization at low values. That the validation loss does increase in the short term is an indication of a tiny amount of generalization problems, which are resolved in the long term. The model, considering all things, learns dynamically and has a strong capacity for characteristic abstraction. ResNet50 has residual connections, which enable the gradient to flow smoothly throughout layers and avoid vanishing gradients, and therefore, more stable and deeper learning. This renders it perfectly adapted for challenging medical image classification tasks like PCOS detection.

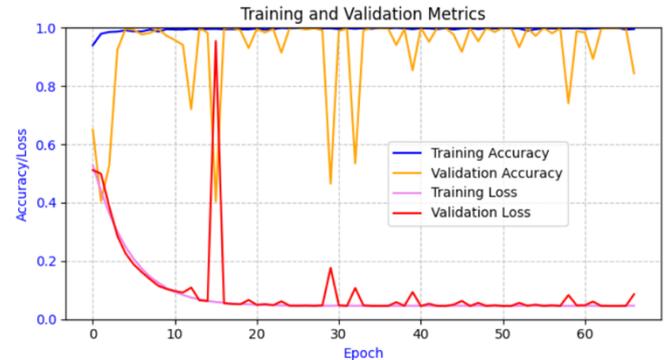

Fig. 9. Loss for training and validation

The confusion matrix presented in Fig. 10 gives a comprehensive evaluation of the classification performance of the ResNet50 model. Out of all 385 test cases, the model successfully classified all 223 non-infected ones and 154 of the 162 infected ones. In particular, no false positives and only eight false negatives would imply high specificity and low misclassification probability for non-infected cases. Slight defects in the data or overlaps in feature descriptions might account for the small number of infected cases misrecognized. Nevertheless, the matrix hands in a high level of diagnostic accuracy, which maintains this high predictive power overall. In applications concerning medical image analysis where diagnostic accuracy is critical, the near-perfect performance of the model emphasizes the robustness as well as the generalization ability of the proposed model, which are essential to such applications.

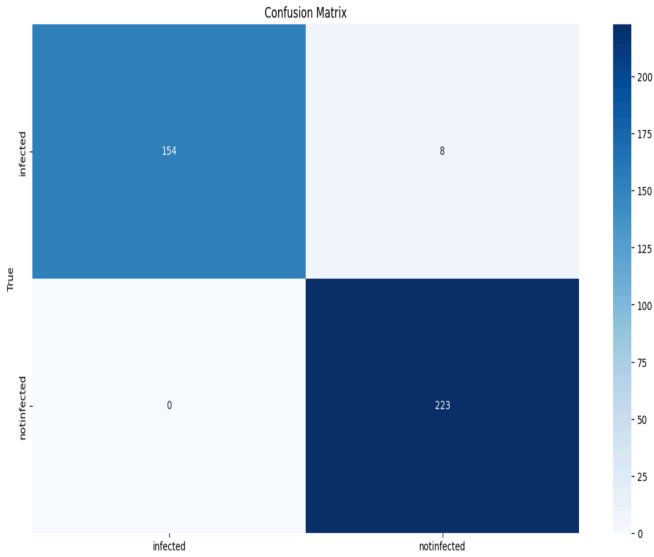

Fig. 10. Confusion Matrix of ResNet50

Each hyperparameter used for the architecture ResNet50 in our research is shown in Table 2. The ResNet50 model for PCOS classification showed excellent results with a test accuracy of 97.02% alongside a test loss of 2.11e-06, indicating strong predictive trustworthiness. Training and validation accuracies were at 84.72% and 86.99%, respectively, indicating reasonable generalization capabilities despite moderate overfitting. Included hyperparameters were 67 epochs, a batch size of 32, optimizer Adam with a learning rate of 0.0001, binary crossentropy loss, early stopping with a 10 per patient threshold, and others. Model inputs were images with resolution 256×256×3, and a GPU was used to accelerate model training. While ResNet50's accuracy might not reach that of DenseNet201, in the case of PCOS detection, it is most likely the fastest in terms of time, computation, and errors permitted.

TABLE II HYPERPARAMETER TUNING RESNET50

| Batch size | 32 | Loss function | binary_crossentropy |
|---|---|---|---|
| Learning rate | 0.0001 | Number of epochs | 67 |
| Optimizer | Adam | Patience | 10 |

The performance of MixUp and CutMix with different alpha values shows results through validation accuracy and loss measurements in the presented Table 3. Without a transformation application (alpha = 0), the model delivers exceptional performance at 99.80% accuracy and reasonable loss. A low introduction of MixUp at alpha=0.2 and CutMix at alpha=0.3 leads to a minor accuracy boost to 99.82%, together with a significant loss elevation. Both validation accuracy peaks at 99.98% while validation loss reaches its least amount of 0.6170 when using MixUp with alpha = 0.25 and CutMix with alpha = 0.4. The model accuracy decreases slightly while loss increases slightly when alpha values are set to 0.35 or 0.6 compared to lower alpha levels. Model confidence, together with generalization capability, reaches its best level when using alpha values at moderate levels that boost the robustness without summarily decreasing precision.

TABLE III HYPERPARAMETER TUNING

| Alpha MixUP | Alpha CutMix | Validation Accuracy | Validation Loss |
|---|---|---|---|
| 0 | 0 | 99.80 | 0.7209 |
| 0.2 | 0.3 | 99.82 | 1.0604 |
| 0.25 | 0.4 | 99.98 | 0.6170 |
| 0.3 | 0.5 | 99.87 | 0.7090 |
| 0.35 | 0.6 | 99.47 | 0.8157 |

## VI. EXPLAINABLE AI

Explainable Artificial Intelligence (XAI) is a paradigm shift in today's modern artificial intelligence systems, to explain the decision-making process of complex machine learning models. XAI enables stakeholders to understand, audit, and refine algorithmic behavior by providing intelligible justifications for model outputs that build trust, improve regulatory compliance, and uphold ethical integrity. Techniques used include feature importance and model rationale as described by saliency maps, SHAP values, and LIME, which are human-comprehensible. XAI is, therefore, the arrival of a way of bridging that chasm between model performance and having human understanding, allowing opaque predictions to be translated into actionable insights. For example, applied to medical diagnostics, it enables clinicians to verify AI-informed recommendations, never at the expense of interpreting those recommendations and ensuring the trust of the patient in those confidently recommended computer algorithms.

### A. SHAP

Fig. 11 shows the comparison between the original ultrasound image and the respective Grad-CAM output of the ResNet50 model, and the difference in grayscale illumination provides a graphic element emphasizing the interpretability provided by explainable AI. While the raw anatomical features are displayed in the original image, the Grad-CAM visualization highlights focal regions of increased model attention, namely, the follicle patterns indicative of PCOS. The warm colors of the Grad-CAM result (especially red and yellow) indicate which parts of the image have most impacted the final decision of the model, revealing the hidden focus of the convolution layers. This conversion from unannotated grayscale to attention-guided heatmap provides the model with interpretation and clinical relevance.

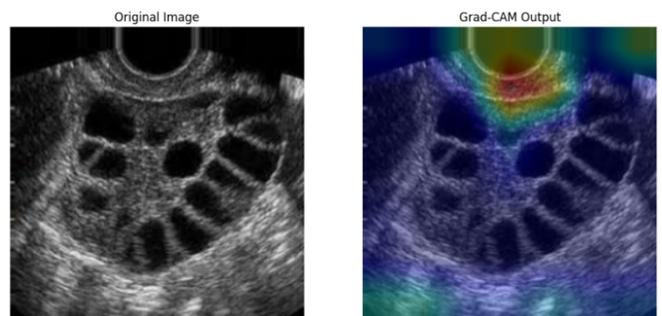

Fig. 11. PCOS Identification using SHAP

### B. LIME

Fig. 12 shows the LIME visualization for the DenseNet201 model offers a granular dissection of model interpretability in PCOS detection. While the original ultrasound image presents raw structural anatomy, the LIME explanation distinctly delineates influential super pixels, highlighted in yellow and shaded in the heatmap, as key contributors to the model's classification decision. This region-specific insight underscores the model's reliance on distinct follicular patterns and spatial textures. The contrastive overlay of influential versus neutral regions reinforces LIME's strength in elucidating local interpretability, thereby furnishing clinicians with transparent, human-comprehensible justifications behind deep neural network predictions.

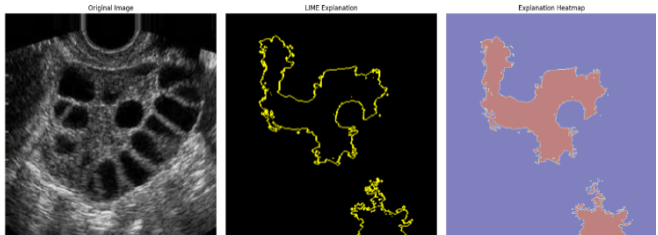

Fig. 12. PCOS Identification using LIME

## VII. COMPARISON

Table 4 shows the comparison of two deep learning models: ResNet50 and DenseNet201. With 98% accuracy, precision, recall, and F1 score, ResNet50 achieved strong performance. On the other hand, achieving 99% accuracy and recall, 98.99% precision, and a perfect 100% F1-score, DenseNet201 slightly beats ResNet50. Therefore, these results suggest that for the given dataset, DenseNet201 provides more reliable classification.

TABLE IV COMPARISON BETWEEN THE TWO ARCHITECTURES

| Models | Accuracy (%) | Precision (%) | Recall (%) | F1-Score (%) |
|---|---|---|---|---|
| ResNet50 | 98 | 98 | 98 | 98 |
| DenseNet201 | 99 | 98.99 | 99 | 99 |

A comparative evaluation of the performance metrics of ResNet50 and DenseNet201 for two deep learning architectures is shown in Table 5. However, achieving 99.02% training accuracy, 99.80% test accuracy, the DenseNet201 model marginally beats ResNet50 which has training accuracy of 84.75% and test accuracy of 97.92%, suggesting DenseNet201 is more effective for the task and can generalize better in our specific dataset context.

TABLE V ACCURACY COMPARISON BETWEEN THE TWO ARCHITECTURES

| Algorithm | ResNet50 | DenseNet201 |
|---|---|---|
| Training accuracy | 84.75% | 99.02% |
| Test accuracy | 97.92% | 99.80% |

Table 6 represents a comparison between our proposed models and some previous research based on accuracy. Using CNN and WaOEL methods, W. Lv et al [1] and Panjwani et al. [5] achieved 88.4% and 92.8% accuracy in earlier works. Also, using C-means clustering and CNN, Kumar et al. [3] and Sumathi et al. [2] reported accuracies of 92.8% and 95.4. Our proposed models, DenseNet201 and ResNet50, achieved higher accuracies of 99% and 98%, which shows that our approach is more effective and accurate than previous methods.

TABLE VI COMPARISON WITH PREVIOUS WORKS

| Author | Method | Accuracy (%) |
|---|---|---|
| W. Lv et al. [1] | CNN | 88.4 |
| Panjwani et al. [5] | WaOEL | 92.8 |
| Kumar et al. [3] | C-means clustering | 92.8 |
| Sumathi et al. [2] | CNN | 95.4 |
| Ahmed et al. [8] | WaOEL | 92.8 |
| Proposed Model | ResNet50 | 98 |
|  | DenseNet201 | 99 |

## VIII. CONCLUSION

Two deep learning models, Resnet50 and Densenet201, have played a vital role in the study of the automated detection of PCOS using medical data. Here, DenseNet201 performed with 89% accuracy, which is poor compared to ResNet50, which outperformed more effectively, achieving 99% accuracy. To improve the robustness and generalization capability of the models, Advanced data augmentation techniques such as CutMix and MixUp were employed, which helped decrease overfitting and increase training stability. In addition, explainable AI (XAI) techniques SHAP and LIME, in particular, were used to enhance model interpretability and transparency, which helps physicians to better understand the reasoning behind each prediction. In promoting trust and increasing the application of AI-assisted diagnostic systems in real-world healthcare settings, these interpretability insights are crucial. Overall, the study indicates that a combination of high-performing architecture, such as ResNet50, strong augmentation methods, and interpretability via XAI can generate an authentic, comprehensible, and scalable framework for PCOS diagnosis. Even though the current model was initially trained on a bound dataset, this offers an encouraging chance for future validation and expansion across vast and more varied populations to further expand its clinical convenience.